\documentclass[fleqn,usenatbib]{mnras}

\usepackage[T1]{fontenc}

\DeclareRobustCommand{\VAN}[3]{#2}
\let\VANthebibliography\thebibliography
\def\thebibliography{\DeclareRobustCommand{\VAN}[3]{##3}\VANthebibliography}

\usepackage{graphicx}	% Including figure files
\usepackage{amsmath}	% Advanced maths commands
\usepackage{amssymb}	% Extra maths symbols

\usepackage{newtxtext,newtxmath}

\usepackage{xcolor}

\title[Method to observe jovian radio emission with multiple LOFAR stations at high-resolution]{Observing Jupiter's radio emissions using multiple LOFAR stations: a first case study of the Io-decametric emission using the Irish IE613, French FR606 and German DE604 stations}

\author[C. K. Louis et al.]{
Corentin. K. Louis$^{1}$\thanks{E-mail: corentin.louis@dias.ie},
C. M. Jackman$^{1}$,
J.-M. Grie\ss meier$^{2,3}$,
O. Wucknitz$^{4}$,
D. J. McKenna$^{1,5}$,
\newauthor
P. C. Murphy$^{5,1}$,
P. T. Gallagher$^{1,5}$,
E. Carley$^{1}$,
D. \'O Fionnagáin$^{6}$,
A. Golden$^{6,7}$,
J. McCauley$^{5}$,
\newauthor
P. Callanan$^{8}$,
M. Redman$^{9}$,
C. Vocks$^{10}$
\\
$^{1}$School of Cosmic Physics, DIAS Dunsink Observatory, Dublin Institute for Advanced Studies, Dublin 15, Ireland,\\
$^{2}$LPC2E - Université d’Orléans /CNRS, France,\\
$^{3}$Station de Radioastronomie de Nançay, Observatoire de Paris, PSL Research University, CNRS, Univ. Orléans, 18330 Nançay, France,\\
$^{4}$Max-Planck-Institut für Radioastronomie, Auf dem Hügel 69, 53121 Bonn, Germany,\\
$^{5}$School of Physics, Trinity College Dublin, Dublin 2, Ireland,\\
$^{6}$Astrophysics Research Group, School of Mathematics, Statistics and Applied Mathematics, National University of Ireland Galway,
University Road, Galway, Ireland,\\
$^{7}$ Armagh Observatory and Planetarium, College Hill, Armagh, N. Ireland,\\
$^{8}$Department of Physics, University College Cork, Cork, Ireland,\\
$^{9}$Centre for Astronomy, School of Physics, National University of Ireland Galway, University Road, Galway, Ireland,\\
$^{10}$Leibniz-Institut für Astrophysik Potsdam (AIP), An der Sternwarte 16, 14482 Potsdam, Germany.
}

\date{Accepted XXX. Received YYY; in original form ZZZ}

\pubyear{2021}

\begin{document}
\label{firstpage}
\pagerange{\pageref{firstpage}--\pageref{lastpage}}
\maketitle

\begin{abstract}
The Low Frequency Array (LOFAR) is an international radio telescope array, consisting of 38 stations in the Netherlands and 14 international stations spread over Europe. Here we present an observation method to study the jovian decametric radio emissions from several LOFAR stations (here DE604, FR606 and IE613), at high temporal and spectral resolution. This method is based on prediction tools, such as radio emission simulations and probability maps, and data processing. We report an observation of Io-induced decametric emission from June 2021, and a first case study of the substructures that compose the macroscopic emissions (called millisecond bursts). The study of these bursts make it possible to determine the electron populations at the origin of these emissions. We then present several possible future avenues for study based on these observations. The methodology and study perspectives described in this paper can be applied to new observations of jovian radio emissions induced by Io, but also by Ganymede or Europa, or jovian auroral radio emissions.

\end{abstract}


%%%%%%%%%%%%%%%%%%%%%%%%%%%%%%%%%%%%%%%%%%%%%%%%%%

%%%%%%%%%%%%%%%%% BODY OF PAPER %%%%%%%%%%%%%%%%%%

\section{Introduction}

In our solar system, Jupiter is the planet with the most intense radio emissions, covering the broadest frequency range from a few kHz (where Quasi-Periodic bursts are seen) up to $40$~MHz (the decametric, or DAM, emissions). From the ground, only the DAM emissions are observable, as all other lower frequency radio emissions are blocked by the ionospheric cut-off at $\sim10$~MHz. These DAM radio emissions were discovered quite early by \citet{1955JGR....60..213B}. A few years later, \citet{1964Natur.203.1008B} discovered that part of the decametric emission was controlled by the interaction between Jupiter's magnetosphere and the Galilean moon Io. 
%Therefore these decametric emissions are observed since more than half a century, using ground- and space-based instruments.
The decametric emissions are known to be generated by the electron cyclotron maser instability (CMI) in the jovian magnetosphere, which occurs when a circularly polarised wave resonates with the gyration movement of electrons with relativistic energies \citep{1958AuJPh..11..564T, 1963Nat...198.4875H, 1979ApJ...230..621W, 1998JGR...10320159Z, 2006A&ARv..13..229T, 2017GeoRL..44.4439L, 2020GeoRL..4790021L}. The source regions of the DAM emission are located above the atmosphere, on magnetic field lines of magnetic apex (distance of the magnetic field lines at the magnetic equator) between 15 and 50~R$_\mathrm{J}$ \citep[{1~R$_\mathrm{J} = 71492$~km the Jovian radius,}][]{2019GeoRL..4611606L}. The CMI gives rise to emission at the local electron cyclotron frequency $f_\mathrm{ce}$ (which is proportional to the local magnetic field amplitude B), along a hollow cone  \citep[with a thickness of $\sim 1 \degr$,][]{2000JGR...10516053K} at large angle with respect to the local magnetic field line \citep[from $\sim 75\degr$ up to $90\degr$,][]{2017GeoRL..44.9225L, 1986PhFl...29.2919P}. From the observer's point of view, these emissions have an arc-shape in a time-frequency map \citep{2017A&A...604A..17M}.

Our knowledge on the CMI emissions allows us to simulate these arcs \citep{2019A&A...627A..30L}, which makes it possible to retrieve the energy of the electrons that produce these emissions \citep{2008GeoRL..3513107H, 2017GeoRL..44.9225L}, as well as discover new components, such as decametric emissions induced by the Galilean moons Europa and Ganymede \citep{2017JGRA..122.9228L}.

However, the physics of the substructures that compose these macroscopic emissions is less well known. Their study requires observations at high temporal and spectral resolution. Unfortunately, most instruments do not have sufficiently high resolution to make the required measurements. For instruments on board space missions, the telemetry often requires reduced frequency and time resolution. For ground-based radio telescopes, high spatial resolution requires tied-array beam forming or interferometric measurements with a large number of antennas, but also the ability to store very large amounts of data. The latest generation of radio telescopes, such as LOFAR \citep{2013A&A...556A...2V} or NenuFAR \citep{2012sf2a.conf..687Z, zarka:hal-01196457}, provides measurements at very high temporal and spectral resolution.

 Only a few authors have looked at these millisecond micro-bursts that composed the arc-shape radio emission. \citet{1996GeoRL..23..125Z} were able to observe these millisecond bursts of Io-DAM emission using the Nançay Decameter Array \citep[NDA,][]{1980Icar...43..399B} and the UTR-2 \citep[Ukrainian T-shaped Radio telescope,][]{1978Anten..26....3B} observatory, with a $10$~ms per $13$~kHz resolution, as well as \citet{2007P&SS...55...89H} using only the NDA observatory with a $3$~ms per $50$~kHz resolution. These two studies showed that these millisecond bursts are due to electron bunches propagating along the magnetic field lines. Abrupt changes in the drift rate of these millisecond bursts can also reveal electric potential drops along these field lines. \citet{2007JGRA..11211212H} were able to model these millisecond bursts by assuming that electrons are accelerated by Alfvén waves in the Io flux tube. \citet{2004P&SS...52.1455Z} showed the feasibility to do fast radio imaging of Jupiter's magnetosphere at radio frequency using a unique LOFAR station.

 In this article, we show that the inclusion of the Irish extension of the LOFAR telescope (I-LOFAR) and the use of both simulations and improved probability maps can extend and enhance the observation of these millisecond structures. In the following, we present the methodology of our observations, and show the results of a first observation taken in June 2021.

%\begin{itemize}
%    \item Here we show that the inclusion of the \textbf{I}rish extension of the \textbf{LOFAR} telescope (I-LOFAR) can improve the observation of these millisecond structures.
%\end{itemize}
 
%The scientific purposes of the method presenting here are to mesure the jovian decametric radio emissions to put constrains on their characteristics (beaming angle, electron energy, exact position of the sources, ...)
%\begin{itemize}
 %   \item Mesure the jovian decametric radio emissions to put constrains on their characteristics (beaming angle, electron energy, position of the sources,…)
%    \item Use LOFAR to obtain high temporal and spectral resolution in the frequency range [8-40] MHz
%    \item Use different LOFAR stations distributed across Europe to conduct interferometry
%    \item Use the westernmost position of the Irish station to enlarge the east-west spreading
%    \item Some successful observations, that are being analyzed,  have already been obtained during an interferometric project in LOFAR cycle 0 (PIs: O. Wucknitz \& P. Zarka)
%    \item First new observing campaign has been done in December 2020 and June 2021, using the DE604 (Postdam, Germany), FR606 (Nançay, France) and IE613 (Birr Castle, Ireland) stations, in an observing window based on ExPRES simulations
%\end{itemize}

\section{Methodology}

\begin{figure}
\centering
 \centerline{\includegraphics[width=1\linewidth]{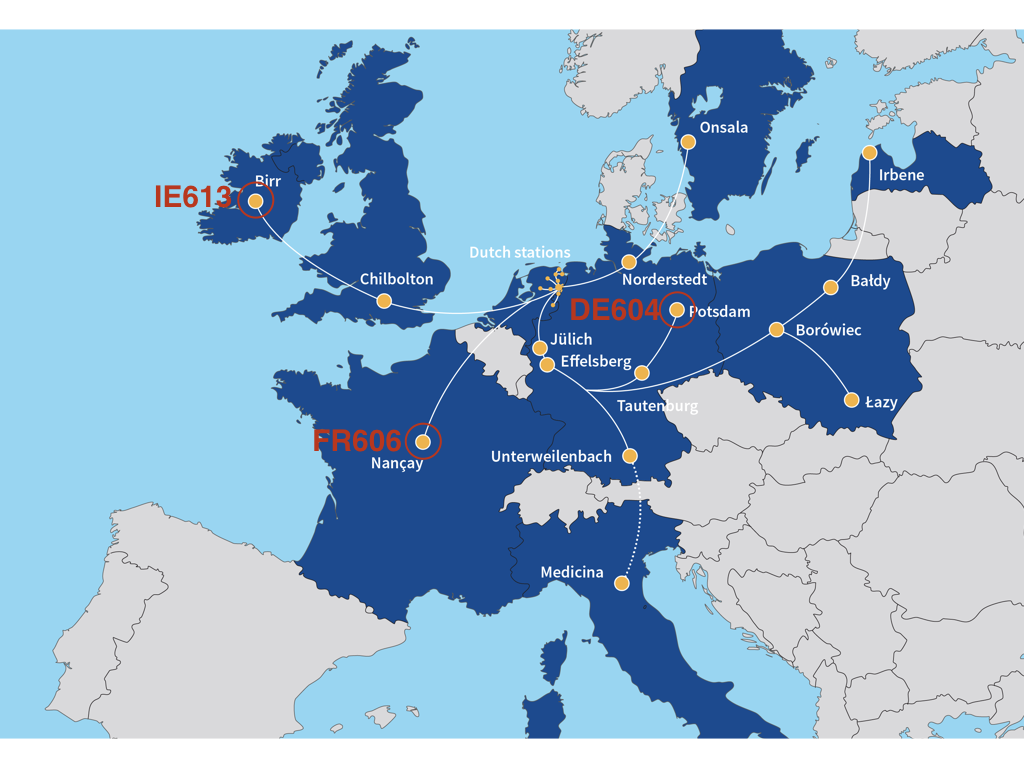}}
\caption{Map showing the location of LOFAR stations across Europe. 38 stations are located in the Netherlands, while 14 (soon 15) stations are located in 7 (soon 8) other European countries. Here are highlighted the three stations that were used in the first case study described in Section \ref{sec:case_study}: the Irish IE613, the French FR606 and the German DE604 stations.}
\label{fig:LOFAR_map}
\end{figure}

\subsection{LOFAR stations}
The Low Frequency Array (LOFAR) is an international radio telescope array (see Figure \ref{fig:LOFAR_map}, consisting of 38 stations in the Netherlands and 14 international stations spread over Europe. In this present study, only the DE604, FR606 and IE613 stations are used (highlighted in red Fig. \ref{fig:LOFAR_map}). 

At each LOFAR station, a backend is used to record and analyse the raw beam formed data in real-time. At the DE604 and FR606 stations, raw data packets were written to disk after lossless compression (using the  \emph{Zstandard} algorithm). At the IE613 station, the REALTA \citep[REAL-time Transient Acquisition,][]{2021A&A...655A..16M} backend was used. Similar successes have already been achieved with these different backends and used to conduct comparative studies \citep[see e.g.,][]{2021AA..654.A43G}. After data has been recorded to disk, \emph{digifil} \citep[{as part of the \emph{dspsr} package,}][]{2011PASA...28.14vanS} was used at all stations to generate the the antenna (auto-)correlation factors X/Y from the underlying voltage data. Finally, a Jovian emission processing pipeline\footnote{\url{https://github.com/CorentinLouis/ILOFAR}} has been run to retrieve the Stokes I (intensity) and V (circular polarization) parameters from the antenna (auto-)correlation factors.

For the study of the jovian decametric radio emissions (extending up to $40$~MHz), only the LOFAR Low Band Antenna (LBA) is used, which allows to make observations in the $8$-$90$~MHz frequency range, with a temporal resolution that can go down to $5$~ns.

%\begin{itemize}
%    \item The Low Frequency Array (LOFAR) is an international radio telescope array, consisting of 38 stations in the Netherlands and 15 international stations spread over Europe.
%    \item I-LOFAR is the Irish complement to this network and the $13^\mathrm{th}$ international station to be built in Europe.
%    \item It allows Irish astrophysical research to be integrated with one of the most sophisticated telescopes on the planet. 
%    \item LOFAR Low Band Antenna is able to make observations in the $8$ to $90$~MHz frequency range. For Jupiter's observation, only the lower part of this range is interesting (from $8$ to $40$~MHz)
%    \item We can go down to a $5$~$\mu\mathrm{s}$ per $12.2$~kHz resolution
%\end{itemize}

\subsection{Optimal observation conditions}

To be able to observe Jupiter's radio emissions in the best possible conditions, the ionosphere should be in a quiet state to avoid supplementary interference, thus during night. It is therefore optimal to observe when Jupiter is near opposition. Jupiter also has to be high enough in the sky ($\gtrsim 10 \degr$), to get more signal in the lobe of the antennas. A good approximation of the international LOFAR antennas effective area can be given by the following equation \citep{2021AA..654.A43G}:
\begin{equation}
    A_\mathrm{eff} = A_{\mathrm{eff}_\mathrm{max}} ~ \mathrm{cos}^2 z, 
\end{equation}
where $z$ is the zenith angle of the source and $A_{\mathrm{eff}_\mathrm{max}}$ the maximal frequency-dependent effective area \citep[{$A_{\mathrm{eff}_\mathrm{max}}$ = [$3974.0$-$2516.0$]~m$^2$ in the range [$15$-$30$]~MHz, see Appendix B of}][for more details]{2013A&A...556A...2V}.
%\begin{itemize}
%    \item ionosphere should be in a quiet state, thus it's better to observe during night, when Jupiter is near opposition.
%    \item Jupiter has to be high enough in the sky, to get more signal in the lobe of the antennas \tbd{find a rough estimate of the sensitivity loss as a function of the target's elevation above the horizon, i.e. as a function of the target's position in the antenna lobe}
%    \item 
%\end{itemize}

\subsection{Emission probability}

\begin{figure*}
 \includegraphics[width=\linewidth]{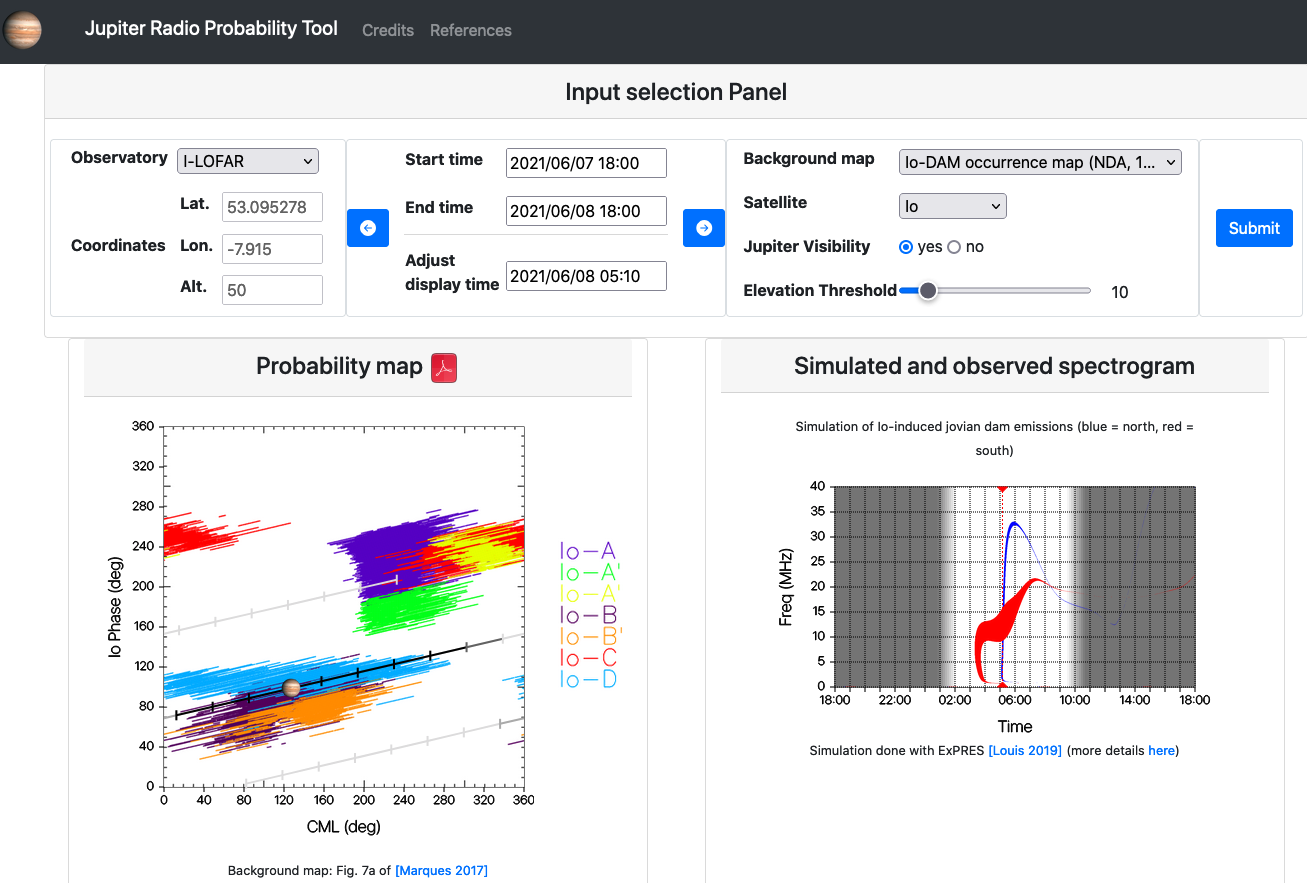}
\caption{Example of results given by the online jovian radio probability tool (\url{https://jupiter-probability-tool.obspm.fr/}).
In this example, the station I-LOFAR has been chosen (\emph{Observatory} entry, upper-left), for an observation between 7 June 2021 18:00 UTC and 8 June 2021 18:00 UTC (\emph{Start Time} and \emph{End Time} entries, upper-middle). The left-hand panel displays the occurrence probability map of Io-DAM emission (\emph{Background map} entry, upper-right), in function of the Galilean moon position with respect to the observer (phase of the moon, y-axis) and the position of the observer (Central Meridian Longitude, or CML, x-axis). The different colors correspond to different Io-DAM emissions (see text). The position of Jupiter in this map for the chosen time window is given by the black line (one-hour tick). The miniature image of Jupiter gives the exact position for the time given in the \emph{Adjust displayed time} entry (upper-middle). The right-hand panel displays the ExPRES simulation for a given moon (\emph{Satellite} entry, upper-right). Blue corresponds to emission sources in the northern hemisphere while red corresponds to emission sources in the southern hemisphere. The thin red dashed line corresponds to the time given in the \emph{Adjust displayed time} entry (upper-middle). The white area gives the time window in which Jupiter is visible above a certain elevation (\emph{Elevation Threshold} entry, upper-right).}
\label{fig:proba_tool}
\end{figure*}

Observations at very high temporal resolutions are very demanding in terms of disk space ($\sim 7.6$~Gb per minute of observation in the case of this study). We must therefore be sure that Jovian radio emission will be observable during the time window when Jupiter is visible in the sky (at most 8 hours per day, due to the rotation of the Earth). 

To predict jovian radio emissions, we use probability maps and simulations, that have been both gathered in an interactive online tool, called \emph{Jupiter Radio Probability Tool}\footnote{\url{https://jupiter-probability-tool.obspm.fr/}\label{ft:probatool}}. An example is displayed in Figure \ref{fig:proba_tool}. This tool allows the user to choose (i) an observatory site from a pre-established list (\emph{Observatory entry}, upper-left), or to enter the GPS coordinates (here `I-LOFAR' is chosen as the Observatory), (ii) a time window (\emph{Start Time} and \emph{End Time} entries, upper-middle), here from 7 June 2021 18:00 UTC to 8 June 2021 18:00 UTC), and finally (iii) a background map (upper-right entry) for the probability map (left-panel).

The probability maps (see Fig. \ref{fig:proba_tool} left panel) show the position of the visible emissions as a function of the jovian longitude of the observer (Central Meridian Longitude, CML) and the phase of the Galilean moon (position of the satellite in the observer's frame, counted positively in the direction of rotation of the satellite, with the origin at the opposition). The example Figure \ref{fig:proba_tool} (left panel) displays the probability map for Io-DAM emission. The different colors corresponds to different type of Io-DAM emission, corresponding to radio sources at different positions with respect to the observer \citep[A: North-East of Jupiter, B: North-West, C: South-East, D: South-West; see e.g., Fig. 2 of ][]{2017A&A...604A..17M}. The position of the observer in these maps is indicated by the position of the miniature image of Jupiter, with the black line displaying the location during the chosen time window (each tick is one hour). Several probability maps have been published to predict auroral or Galilean moons induced decametric emissions such as \citet[][the one displayed here]{2017A&A...604A..17M}, \citet{2018A&A...618A..84Z, 1993A&AS...98..529L, 2017JGRA..122.9228L, 2017pre8.conf...45Z}.

We also simulate the radio emission (see Fig. \ref{fig:proba_tool} right panel), by using the ExPRES tool \citep[Exoplanetary and Planetary Radio Emission Simulator,][]{2019A&A...627A..30L}. This code is here used to produce simulations of the jovian radio emission linked to the interaction between Jupiter and the Galilean moons \citep[Io, Europa, Ganymede, see ][for the database]{2020ExPRES_simulations_data_collection}, based on the Cyclotron Maser Instability equations. More precisely, the simulations in this database were produced using as input parameters the JRM09 magnetic field model \citep[based on in situ measurements from the Juno mission, ]{2018GeoRL..45.2590C} and the \citet{1981JGR....86.8370C} current sheet model to reproduce the magnetic field lines of Jupiter's magnetosphere, a $3$~keV electron energy and a loss cone electron distribution function. Using these standard parameters, the uncertainty for Io-DAM emission is expected to be well within a $2$~hours window \citep{2017pre8.conf...59L} around the predicted arc. The example of Fig. \ref{fig:proba_tool} (right panel) displays two arc-shape emissions related to Io-Jupiter interaction: the blue one is a northern emission (corresponding to an Io-B emission) while the red one is a southern emission (Io-D emission). The white area corresponds to the time when Jupiter is visible above an adjustable elevation (here above $10 \degr$), while the grey-shaded areas correspond to the time when Jupiter is below this elevation. The observer position indication, given in the left-hand side panel (with the miniature image of Jupiter), is here represented by the thin vertical red dashed line (with the corresponding time in the \emph{Adjust display time} entry, upper-middle).

Using this online tool, we are thus able to determine when the probability of observing Io-, Europa-, Ganymede- or auroral decametric emissions is the highest.

%\begin{itemize}
%    \item Observations at very high temporal resolutions are very demanding in terms of disk space, we must therefore be sure that we will have visible jovian radio %emission
%    \item For that, we use different tool:
%    \begin{itemize}
%        \item first one is the ExPRES tool \citep<Exoplanetary and Planetary Radio Emission Simulator>[]{2019A&A...627A..30L}. This code allows us to produce %simulations of the emissions linked to the interaction between Jupiter and the %Galilean moons (Io, Europa, Ganymede) %\citep{2020ExPRES_simulations_data_collection}
%        \item the second one is probability map that have been published by different %authors, such as \citep{2017A&A...604A..17M, 2018A&A...618A..84Z, %1993A&AS...98..529L, 2017JGRA..122.9228L, 2017pre8.conf...45Z}
%        \item For that, we use the Jupiter Probability Tool %\url{https://jupiter-probability-tool.obspm.fr/} which gather both at the same %place (see Figure \ref{fig:proba_tool}).
%    \end{itemize}
%\end{itemize}

\section{A first case study: observation of an Io-decametric emission}
\label{sec:case_study}

\begin{figure*}
 \includegraphics[width=0.88\textwidth]{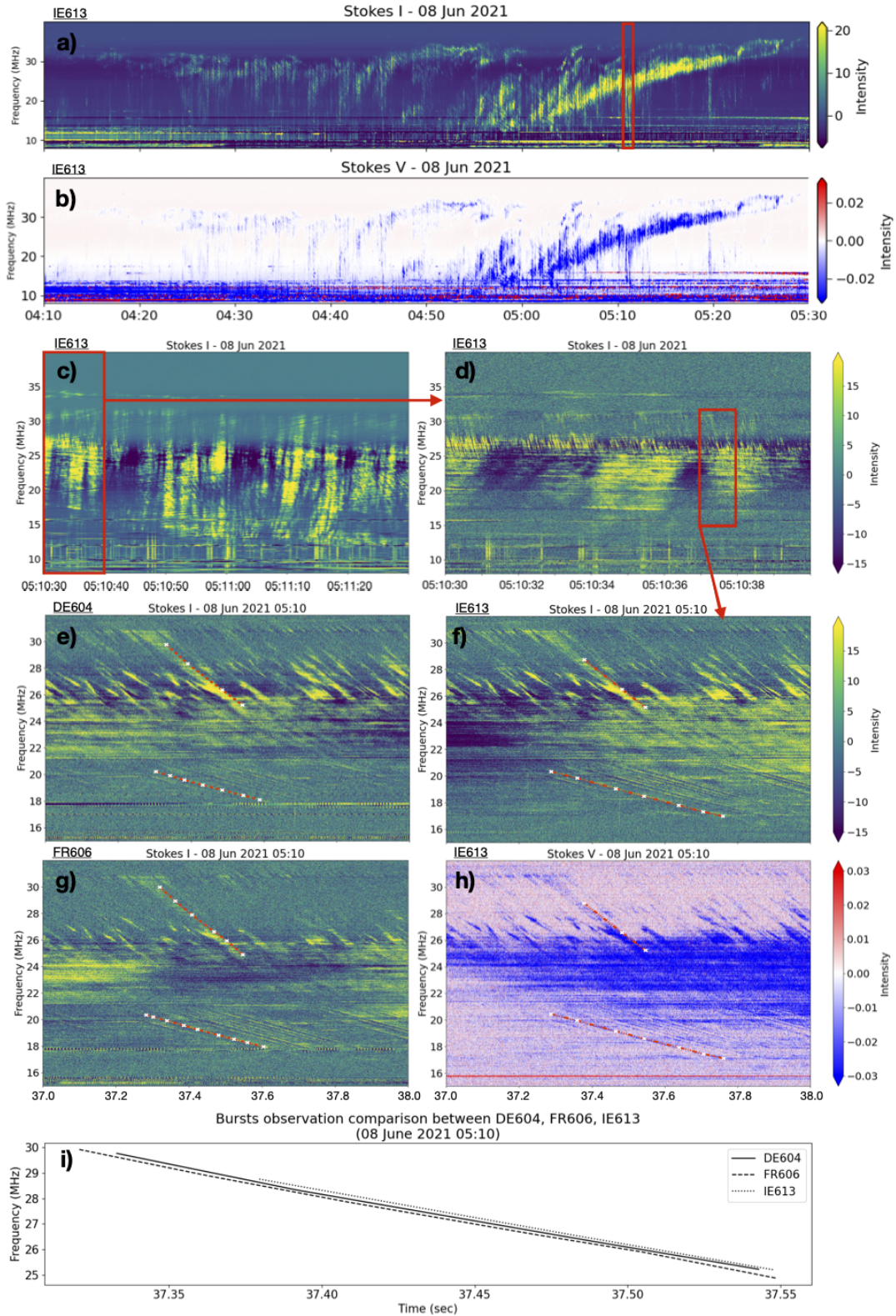}
\caption{Observation of an Io-decametric emission, displayed in a time-frequency map (called \emph{dynamic spectrum}), on different time windows, and for the stations IE613 (panels a-d,f,h), DE604 (panel e) and FR606 (panel g). Panels (a) and (b) respectively display Stokes I (corresponding to the intensity of the emission) and Stokes V (corresponding to the degree of circular polarization) parameters between 04:10 and 05:30 UTC on 8 June 2021. Panels (c) and (d) are two zoom-in inside the red boxes, between (c) 05:10:30 and 05:11:30, and (d) 05:10:30 and 05:10:40, showing the data at an higher-resolution. Panel (f) is a zoom of panel (d) of 1 second, between 05:10:37 and 05:10:38, in the range [$15$-$32$]~MHz, at the highest resolution use in this study, i.e. $81.92$~$\mu \mathrm{s}$ per $12.2$~kHz. Panels (e) and (g) are the equivalent of panel (f), displaying the data from the DE604 and FR606 stations, respectively. Panel (h) is the equivalent of panel (f), displaying the Stokes V parameter (degrees of circular polarization).
Panels (c) to (g) share the same intensity colour bar (displayed to the right of panels (d) and (f)).
In panels (e) to (h) the red lines show example of drifting millisecond bursts. Panel (i) shows a comparison between the highest frequency bursts observed by the three stations (highlighted in panels (e-h).}
\label{fig:obs_IE613}
\end{figure*}

As a first case study, and based on the prediction shown in Figure \ref{fig:proba_tool}, we observed Jupiter pre-dawn on 8 June 2021, between 04:00 and 07:00 UTC, with three different LOFAR stations: IE613 in Ireland, FR606 in France and DE604 in Germany (see Fig. \ref{fig:LOFAR_map} for their respective position). In this study the data were processed to obtain a resolution of $81.92$~$\mu \mathrm{s}$ per $12.2$~kHz, and only the [$8$-$40$]~MHz frequency range is displayed ($40$~MHz being the upper limit of the Jovian radio emissions).

As predicted by the prediction tool (see Figure \ref{fig:proba_tool}), we observe (Figs. \ref{fig:obs_IE613}a,b, IE613 data) an emission produced by the Io-Jupiter interaction. The emission has a vertex early shape (shape of an `opening parentheses') and displays a negative circular (Right-Handed) polarization (Fig. \ref{fig:obs_IE613}b), which gives the information that this emission is an Io-B emission (coming from the North-West side of Jupiter as seen from the observer point of view, thus from the North-dawn side of Jupiter as seen from Earth). The main emission (emitted from the flux tube connected to Io) is observed from  $\sim$ 05:00 to 05:30, while emission observed just before is produced in flux tubes connected to the tail upstream of Io (called secondary emission).

Figure \ref{fig:obs_IE613}c displays a one minute zoom of Figure \ref{fig:obs_IE613}a between 05:10:30 and 05:11:30. In this dynamic spectrum, we can see that the general arc shape of this emission has a detailed substructure. If we zoom in even further, in a 10 seconds window (Figure \ref{fig:obs_IE613}d), we then can clearly see that the Io emission is composed of many millisecond bursts, that present a negative drift (decreasing frequency with time). The slope of the millisecond bursts gives a direct indication of the direction of propagation of the electrons responsible for these emissions. Indeed, the Cyclotron Maser Instability produces emission at a frequency close to the electron cyclotron frequency $f_\mathrm{ce}$, which is directly proportional to the magnetic field amplitude. Because the magnetic field strength decreases with altitude, we can therefore determine from the negative slope  of these bursts (the emission frequency decreasing with time) that the electrons are propagating upward (from the planet to the higher altitudes).

Figures \ref{fig:obs_IE613}f and \ref{fig:obs_IE613}h show a 1 second zoom in (between 05:10:37 and 05:10:38) in the [$15$-$32$]~MHz frequency range of both Stokes I (intensity, panel f) and V (circular polarization, panel h) parameters. Superimposed on these two one-second dynamic spectra are two examples of drifting millisecond bursts (orange lines), at two different frequency ranges. 

The first drifting millisecond burst that is highlighted extends from $30.83$~MHz to $22.32$~MHz, from 05:10:37.274 to 05:10:37.673, thus displaying a drift of $\sim-21.34$~MHz/s. At these frequencies, using the JRM09 magnetic field model \citep[based on in situ measurements of the magnetometer onboard Juno,][that allow to calculate the Jupiter's magnetic field lines]{2018GeoRL..45.2590C}, we can determine than a difference of 1 MHz corresponds to a distance of about $\sim 1000$~km. Using the  $E = 1/2 \mathrm{m}_\mathrm{e} \mathrm{v}^2$ formula, we can determine that this drift corresponds to electrons with an energy of $\sim 1.3$~keV.

The second drifting millisecond burst that is highlighted extends from $20.37$~MHz to $17.04$~MHz, from 05:10:37.289 to 05:10:37.761, thus having a drift of $-7.06$~MHz/s. This corresponds to electrons with an energy of $\sim 0.14$~keV.

Figures \ref{fig:obs_IE613}e,g display the observation acquired by the DE604 and FR606 stations during the same $10$~s displayed Fig. \ref{fig:obs_IE613}f. The same bursts are highlighted. 

Fig. \ref{fig:obs_IE613}i shows a zoom on the bursts at the highest frequencies, observed by the three stations (DE604 as a solid line, FR606 as a dashed line and IE613 as a dotted line). A time shift of $\sim 8$~ms is measured here, with FR606 seeing the burst before DE604 and then IE613. At that time (8 June 2021 05:10), Jupiter was located at $157 \degr 33' 02.6"$ in azimuth and $22\degr 27' 26.9"$ in altitude in the sky. For an European observer, it means than Jupiter was located on the South section in the sky. Therefore in that case, it's more the South/North spreading of the three stations that matters (see Fig. \ref{fig:LOFAR_map}). It is then expected that a radio emission from Jupiter will be observed first by FR606, then by DE604 and finally by IE613.

\section{Perspectives}

The observing method described in this article for observing Jupiter's radio emissions at high resolution could not be more timely. With the Juno mission we have in situ measurements for up to 2025, followed by the arrival of the Europa Clipper and JUICE missions in the late 2020s-early 2030s. The medium-resolution ground support has already existed for many years \citep[with daily observation, e.g.,]{2017A&A...604A..17M}, and the addition of regular high-resolution measurements will allow comparative in situ and remote measurements. These results will be compared to simultaneous observation of the Ultraviolet (UV) emission on Jupiter's atmosphere \citep[by the Space Telescope Imaging Spectrograph instrument on-board the Hubble Space Telescope or by the UV spectrograph on-board the Juno mission,][]{2017SSRv..213..447G}, or to X-ray auroral observation (using e.g., Chandra and XMM). This once-in-a-generation combination of high fidelity in situ and remote sensing measurement will allow us to make breakthrough on several key scientific questions for Jupiter.

We have demonstrated the predictive power of the ExPRES simulations and the probability maps (gathered on the Jupiter Radio Probability Tool), the observing capability of the LOFAR network, and the data analysis methods enabled by software support \citep[e.g., REALTA][]{2021A&A...655A..16M} and domain knowledge (e.g., the search for millisecond bursts in the high resolution data stream). We now highlight several future avenues for the exploitation of this rich LOFAR dataset.

%Due to the longitude/latitude spreading of the different stations, we should be able to observe a statistical time shift between observations of the same drifting millisecond bursts event, especially at the starting and ending edges of the main emission. Once this shift is known, this should allow us to deduce the thickness of the emission cone sheet and the size of the 

The millisecond bursts which we have presented here have the potential to reveal important information about the nature of electron acceleration regions at Jupiter. Specifically, examination of the drifting of the bursts can tell us about the electron population energy that gives rise to this emission, in particular on the differences in electron energies as a function of frequency, but also by comparing the energies of the electrons in the main (linked to Io itself) and secondary (linked to Io's tail) emissions. Particular attention will be paid to abrupt changes in the drift rate of the millisecond bursts, which will reveal electric potential drops along these field lines, and thus acceleration regions of the electrons. In order to conduct robust statistical examination of these bursts with future larger LOFAR observations, we plan to apply automatic detection algorithms \citep[e.g.,][]{2007P&SS...55...89H} which will enable examination of the bursts over an extensive frequency range.

Secondly, we plan to unlock the full power of the radio emissions as a tool to provide information about the nature of the radio source locations themselves. We plan to do this by combining the latest jovian magnetic field \citep[JRM09,][]{2018GeoRL..45.2590C} and current sheet \citep{2018GeoRL..45.2590C} models, with sophisticated inter-station interferometry. This will enable us to estimate the location of sources in longitude, latitude and altitude. When this information is combined with in situ Juno measurements, it will represent an extremely powerful probe of the response of the jovian radio sources to magnetospheric dynamics.

Thirdly, we plan to use the radio emissions to reveal the electron density inside the Io torus. In order to do this we will utilise the property that the CMI perturbation is moving at the Alfven speed $v_A = B / \sqrt{\mu_0 \rho}$ (with $B$ the local magnetic field amplitude and $\rho$ the local electron density), and that Io is surrounded by a dense plasma torus (produced by its volcanic activity which releases about one tonne of plasma per second). This leads to a difference in longitude (called \emph{lead angle}) between the Io instantaneous flux tube (IFT, connected to Io) and the Io active flux tube (AFT, where the sources of the main emission are). Once the exact locations of the sources are known, we will be able to obtain the lead angle in longitude between the IFT and the AFT. Knowing this will give the possibility to estimate the electron density inside the Io torus. Knowing the longitude of the AFT position will make it possible to determine the beaming angle of the emitting cone, as well as determine the electron energy in a second way (the first being using the drift rate of the millisecond bursts), by using the theoretical formula of the cyclotron maser instability 
\begin{equation}
\theta = \mathrm{arccos} \left( \frac{v/c}
{
    \sqrt{
            1-\omega_\mathrm{ce}/\omega_\mathrm{ce_\mathrm{max}}
        }
}\right),
\end{equation}
with $v$ the electron velocity, $\omega_{ce} = 2 \pi f_\mathrm{ce}$ the local electron cyclotron gyration frequency and $\omega_{ce_{max}}$ the maximal electron cyclotron gyration frequency at the footprint of the magnetic field lines.

All these studies are beyond the scope of this paper, and will be the focus of a future scientific work. The methodology and study perspectives described here can later be applied to new observations of jovian radio emissions induced by Io, but also by Ganymede or Europa, or jovian auroral radio emissions.

\section*{Acknowledgements}

This paper is using data obtained with the DE604, FR606 and IE613 stations of the International LOFAR Telescope, constructed by ASTRON, during station-owners time. 
German DE604 station is funded by the Leibniz-Institut für Astrophysik (Potsdam). These observations were carried out in the stand-alone GLOW mode (German LOng-Wavelength array), which is technically operated and supported by the Max-Planck-Institut für Radioastronomie, the Forschungszentrum Jülich and Bielefeld University.
Nançay Radio Observatory FR606 station is operated by Paris Observatory, associated with the French Centre National de la Recherche Scientifique and Université d’Orléans.
Irish IE613 station received funding from Science Foundation Ireland (SFI), the Department of Jobs Enterprise and Innovation (DJEI). The REALTA backend used at IE613 station is funded by SFI and Breakthrough Listen. The Irish-LOFAR consortium consists of Trinity College Dublin, University College Dublin, Athlone Institute for Technology, Armagh Observatory and Planetarium (supported through funding from the Department for Communities of the N. Ireland Executive), Dublin City University, Dublin Institute for Advanced Studies, National University of Ireland Galway and University College Cork.
The authors thank S. Aicardi for developing \url{https://jupiter-probability-tool.obspm.fr}.
C. K. Louis' and C. M. Jackman's work at DIAS is supported by the Science Foundation Ireland Grant 18/FRL/6199.
P.C.Murphy and D.McKenna are supported by Government of Ireland Studentships from the Irish Research Council (IRC). 
D. Ó Fionnagáin is supported by a Government of Ireland Postdoctoral Fellowship from the IRC (GOIPD/2020/145).
We acknowledge support and operation of the GLOW network, computing and storage facilities by the FZ-Jülich, the MPIfR and Bielefeld University and financial support from BMBF D-LOFAR III (grant 05A14PBA) and D-LOFAR IV (grant 05A17PBA), and by the states of Nordrhein-Westfalia and Hamburg.

%%%%%%%%%%%%%%%%%%%%%%%%%%%%%%%%%%%%%%%%%%%%%%%%%%
\section*{Data Availability}
The data can be access upon request to the authors.



\bibliographystyle{mnras}
\bibliography{biblio} % if your bibtex file is called example.bib

%\begin{thebibliography}{99}
%\bibitem[\protect\citetuthoryear{Author}{2012}]{Author2012}
%\bibitem[\protect\citetuthoryear{Others}{2013}]{Others2013}
%\end{thebibliography}

%%%%%%%%%%%%%%%%%%%%%%%%%%%%%%%%%%%%%%%%%%%%%%%%%%

%%%%%%%%%%%%%%%%% APPENDICES %%%%%%%%%%%%%%%%%%%%%

%%%%%%%%%%%%%%%%%%%%%%%%%%%%%%%%%%%%%%%%%%%%%%%%%%

% Don't change these lines
\bsp	% typesetting comment
\label{lastpage}
\end{document}